# A Schmid-Leiman based transformation resulting in perfect inter-correlations of three types of factor score predictors

André Beauducel[1]

November, 1st, 2015

## Abstract

Factor score predictors are to be computed when the individual scores on the factors are of interest. Conditions for a perfect inter-correlation of the regression/best linear factor score predictor, the best linear conditionally unbiased predictor, and the determinant best linear correlation-preserving predictor are presented. When these three types of factor score predictors are perfectly correlated for corresponding factors, the factor score predictors computed from one method will have the virtues of the factor score predictors computed from the other methods. A Schmid-Leiman based transformation for which the three types of factor score predictors are perfectly correlated for corresponding orthogonal factors is proposed.

[1] Dr. André Beauducel, Institute of Psychology, University of Bonn, Kaiser-Karl-Ring 9, 53111 Bonn, Germany, email: beauducel@uni-bonn.de

## Introduction

Since factor scores are not determinate (Guttman, 1955), they cannot be unambiguously computed. However, factor score predictors can be computed as linear combinations of the observed variables in order to represent the individual scores of a latent variable. This might be useful when decisions have to be justified on the individual score level. Several different factor score predictors have meanwhile been proposed (Mulaik, 2010). The properties of different factor score predictors have been investigated by means of simulation studies (Fava & Velicer, 1992) and by means of algebraic considerations (e.g. Beauducel & Hilger, 2015; Krijnen, 2006; Krijnen, Wansbeek & Ten Berge, 1996; McDonald & Burr, 1967; Schneeweiss & Mathes, 1995). According to Grice (2001) and according to Krijnen et al. (1996) there are three main types of factor score predictors: The best linear predictor that is also known as Thurstone's (1935) regression predictor, the conditionally unbiased predictor (Krijnen et al., 1996; Bartlett, 1937), and the correlation-preserving predictor (McDonald, 1981; Ten Berge, Krijnen, Wansbeek & Shapiro, 1999). These three types of factor score predictors represent three desired properties: (a) The best linear predictor has a maximal correlation with the corresponding factor, (b) the conditionally unbiased predictor has zero correlations with non-corresponding factors, and (c) the correlation-preserving predictor has the advantage of preserving the correlations between the factors in the factor score predictor.

McDonald and Burr (1967) have explored the conditions for high correlations between factor score predictors for corresponding factors. They investigated the best linear predictor, a conditionally unbiased predictor, and a correlation preserving predictor. Since the determinant best linear correlation-preserving predictor (Ten Berge, Krijnen, Wansbeek, & Shapiro, 1999) was not available at that time, they explored the Anderson-Rubin's (1956) orthogonal (orthogonality preserving) factor score predictor. They found that the three factor score predictors are perfectly correlated for the one factor model (the Spearman case). Moreover, they describe that the investigated factor score predictors are perfectly correlated in the case of unrotated canonical factor analysis (Rao, 1955). McDonald and Burr (1967) acknowledge that investigators would prefer to use rotated factor loadings, because they can often be interpreted more easily. However, for the rotated factors the correlations between the factor score predictors would generally not be perfect, leading to the problem of choosing the optimal factor score predictor.

To sum up, there are at least three types of factor score predictors corresponding to three different desired properties (Grice, 2001). Moreover, there are conditions for which the correlations between the factor score predictors are one for corresponding factors, so that no choice has to be made (McDonald & Burr, 1967). It can be regarded as a substantial advantage of factor score predictors when they are simultaneously the best linear predictor, conditionally unbiased, as well as correlation preserving. Therefore, the aim of the present paper is (1) to explore further the conditions for perfect correlations between the factor score predictors of corresponding factors and (2) to propose a transformation method based on Schmid-Leiman (1957) that allows to find interpretable factors with perfect correlations between the three different types of factor score predictors.

## Definitions

In order to present the equations defining the three factor score predictors, the definition of the population common factor model is given. The common factor model assumes that $\mathbf{x}$, the random vector of observations of order $p$, is generated by

$$\mathbf{x} = \Lambda\mathbf{f} + \mathbf{e}, \tag{1}$$

where $\mathbf{f}$ is the random vector of factor scores of order $q$, $\mathbf{e}$ the random error vector of order $p$, and $\Lambda$ the factor pattern matrix of order $p$ by $q$. The observations $\mathbf{x}$, the factor scores $\mathbf{f}$, and the error vectors $\mathbf{e}$ are assumed to have an expectation zero ($\varepsilon[\mathbf{x}] = 0$, $\varepsilon[\mathbf{f}] = 0$, $\varepsilon[\mathbf{e}] = 0$). The covariance between the factor scores and the error scores is assumed to be zero ($\mathrm{Cov}[\mathbf{f}, \mathbf{e}] = 0$). The standard deviation of $\mathbf{f}$ is one, the covariance of the observed variables is $\mathbf{xx}' = \Sigma$. The covariance matrix $\Sigma$ can be decomposed by

$$\Sigma = \Lambda\Phi\Lambda' + \Psi^2, \tag{2}$$

where $\Phi$ represents the $q$ by $q$ factor correlation matrix and $\Psi^2$ the $p$ by $p$ covariance matrix of the error scores $\mathbf{e}$ ($\mathrm{Cov}[\mathbf{e}, \mathbf{e}] = \Psi^2$). $\Psi^2$ is assumed to be a diagonal matrix and it will be assumed in this paper that it contains only positive values.

The regression predictor or best linear (BL) predictor is given by $\hat{\mathbf{f}}_{BL} = \mathbf{\Phi}\mathbf{\Lambda}'\mathbf{\Sigma}^{-1}\mathbf{x}$. The condition $\mathbf{B}'\mathbf{\Lambda} = \mathbf{I}$ holds for the class of conditionally unbiased predictors, where $\mathbf{B}$ are the weights for the factor score predictor (Bartlett, 1937). According to Krijnen et al. (1996) the best linear conditionally unbiased (BLCU) predictor is $\hat{\mathbf{f}}_{BCLU} = (\mathbf{\Lambda}'\mathbf{\Sigma}^{-1}\mathbf{\Lambda})^{-1}\mathbf{\Lambda}'\mathbf{\Sigma}^{-1}\mathbf{x}$. Ten Berge et al. (1999) defined a determinant best linear correlation-preserving (DBLCP) predictor, given by $\hat{\mathbf{f}}_{DBLCP} = \mathbf{\Phi}^{1/2}(\mathbf{\Phi}^{1/2}\mathbf{\Lambda}'\mathbf{\Sigma}^{-1}\mathbf{\Lambda}\mathbf{\Phi}^{1/2})^{-1/2}\mathbf{\Phi}^{1/2}\mathbf{\Lambda}'\mathbf{\Sigma}^{-1}\mathbf{x}$. For this predictor symmetric positive (semi) definite matrices are raised to a certain power (e.g. square-root) by raising its eigenvalues to that power. When the power of the eigenvalues is "1/2", this procedure is sometimes called the symmetric square-root (Harman, 1976).

### Conditions for a perfect correlation between $\hat{\mathbf{f}}_{BL}$, $\hat{\mathbf{f}}_{BLCU}$, and $\hat{\mathbf{f}}_{DBLCP}$

Theorem 1 gives a condition for a perfect correlation between $\hat{\mathbf{f}}_{BLCU}$ and $\hat{\mathbf{f}}_{BL}$ for corresponding orthogonal factors.

**Theorem 1**. *If* $\mathbf{\Phi} = \mathbf{I}$ *and* $\mathbf{\Lambda}'\mathbf{\Sigma}^{-1}\mathbf{\Lambda} = \text{diag}(\mathbf{\Lambda}'\mathbf{\Sigma}^{-1}\mathbf{\Lambda})$ *then*

$$\varepsilon\left[\hat{\mathbf{f}}_{BLCU}\hat{\mathbf{f}}_{BL}'\right]\text{diag}(\varepsilon\left[\hat{\mathbf{f}}_{BLCU}\hat{\mathbf{f}}_{BLCU}'\right])^{-1/2}\text{diag}(\varepsilon\left[\hat{\mathbf{f}}_{BL}\hat{\mathbf{f}}_{BL}'\right])^{-1/2} = \mathbf{R}_{BLCU,BL} = \mathbf{I}.$$

*Proof.* The covariance between $\hat{\mathbf{f}}_{BLCU}$ and $\hat{\mathbf{f}}_{BL}$ is

$$\mathbf{C}_{BLCU,BL} = (\mathbf{\Lambda}'\mathbf{\Sigma}^{-1}\mathbf{\Lambda})^{-1}\mathbf{\Lambda}'\mathbf{\Sigma}^{-1}\mathbf{xx}'\mathbf{\Sigma}^{-1}\mathbf{\Lambda}\mathbf{\Phi} = \mathbf{\Phi}. \qquad (3)$$

The correlation between $\hat{\mathbf{f}}_{BLCU}$ and $\hat{\mathbf{f}}_{BL}$ is therefore

$$\mathbf{R}_{BLCU,BL} = \mathbf{\Phi}\,\text{diag}((\mathbf{\Lambda}'\mathbf{\Sigma}^{-1}\mathbf{\Lambda})^{-1})^{-1/2}\,\text{diag}(\mathbf{\Phi}\mathbf{\Lambda}'\mathbf{\Sigma}^{-1}\mathbf{\Lambda}\mathbf{\Phi})^{-1/2}. \qquad (4)$$

The element-wise square-root is calculated for the diagonal elements in Equation 4.

For $\mathbf{\Phi} = \mathbf{I}$ and $\mathbf{\Lambda}'\mathbf{\Sigma}^{-1}\mathbf{\Lambda} = \text{diag}(\mathbf{\Lambda}'\mathbf{\Sigma}^{-1}\mathbf{\Lambda})$ Equation 4 can be transformed into

$$\mathbf{R}_{BLCU,BL} = \text{diag}(\mathbf{\Lambda}'\mathbf{\Sigma}^{-1}\mathbf{\Lambda})^{1/2}\,\text{diag}(\mathbf{\Lambda}'\mathbf{\Sigma}^{-1}\mathbf{\Lambda})^{-1/2} = \mathbf{I}. \qquad (5)$$

This completes the proof.                                                                    □

The same condition is also a basis for a perfect correlation between $\hat{\mathbf{f}}_{DBLCP}$ and $\hat{\mathbf{f}}_{BL}$.

**Theorem 2**. *If* $\mathbf{\Phi} = \mathbf{I}$ *and* $\mathbf{\Lambda}'\mathbf{\Sigma}^{-1}\mathbf{\Lambda} = \text{diag}(\mathbf{\Lambda}'\mathbf{\Sigma}^{-1}\mathbf{\Lambda})$ *then*

$$\varepsilon\left[\hat{\mathbf{f}}_{DBLCP}\hat{\mathbf{f}}_{BL}'\right]\text{diag}(\varepsilon\left[\hat{\mathbf{f}}_{DBLCP}\hat{\mathbf{f}}_{DBLCP}'\right])^{-1/2}\text{diag}(\varepsilon\left[\hat{\mathbf{f}}_{BL}\hat{\mathbf{f}}_{BL}'\right])^{-1/2} = \mathbf{R}_{DBLCP,BL} = \mathbf{I}.$$

*Proof.* The covariance between $\hat{\mathbf{f}}_{DBLCP}$ and $\hat{\mathbf{f}}_{BL}$ is

$$\mathbf{C}_{\mathrm{DBLCP,BL}} = \mathbf{\Phi}^{1/2}(\mathbf{\Phi}^{1/2}\mathbf{\Lambda}'\mathbf{\Sigma}^{-1}\mathbf{\Lambda}\mathbf{\Phi}^{1/2})^{-1/2}\mathbf{\Phi}^{1/2}\mathbf{\Lambda}'\mathbf{\Sigma}^{-1}\mathbf{xx}'\mathbf{\Sigma}^{-1}\mathbf{\Lambda}\mathbf{\Phi}$$

$$= \mathbf{\Phi}^{1/2}(\mathbf{\Phi}^{1/2}\mathbf{\Lambda}'\mathbf{\Sigma}^{-1}\mathbf{\Lambda}\mathbf{\Phi}^{1/2})^{1/2}\mathbf{\Phi}^{1/2}. \tag{6}$$

The corresponding correlation is

$$\mathbf{R}_{\mathrm{DBLCP,BL}} = \mathbf{\Phi}^{1/2}(\mathbf{\Phi}^{1/2}\mathbf{\Lambda}'\mathbf{\Sigma}^{-1}\mathbf{\Lambda}\mathbf{\Phi}^{1/2})^{1/2}\mathbf{\Phi}^{1/2}\mathrm{diag}(\mathbf{\Phi}\mathbf{\Lambda}'\mathbf{\Sigma}^{-1}\mathbf{\Lambda}\mathbf{\Phi})^{-1/2} \ . \tag{7}$$

For $\mathbf{\Phi} = \mathbf{I}$ and $\mathbf{\Lambda}'\mathbf{\Sigma}^{-1}\mathbf{\Lambda} = \mathrm{diag}(\mathbf{\Lambda}'\mathbf{\Sigma}^{-1}\mathbf{\Lambda})$ Equation 7 can be transformed into

$$\mathbf{R}_{\mathrm{DBLCP,BL}} = (\mathbf{\Lambda}'\mathbf{\Sigma}^{-1}\mathbf{\Lambda})^{1/2}(\mathbf{\Lambda}'\mathbf{\Sigma}^{-1}\mathbf{\Lambda})^{-1/2} = \mathbf{I}, \tag{8}$$

since the symmetric square-root and the conventional square-root are identical for diagonal matrices. This completes the proof.                                                          □

Finally, the condition presented in Theorem 1 and 2 is also the basis for a perfect correlation between $\hat{\mathbf{f}}_{\mathrm{BLCU}}$ and $\hat{\mathbf{f}}_{\mathrm{DBLCP}}$ for corresponding orthogonal factors.

**Theorem 3**. *If* $\mathbf{\Phi} = \mathbf{I}$ *and* $\mathbf{\Lambda}'\mathbf{\Sigma}^{-1}\mathbf{\Lambda} = \mathrm{diag}(\mathbf{\Lambda}'\mathbf{\Sigma}^{-1}\mathbf{\Lambda})$ *then*

$$\epsilon\left[\hat{\mathbf{f}}_{\mathrm{BLCU}}\hat{\mathbf{f}}_{\mathrm{DBLCP}}'\right]\mathrm{diag}(\epsilon\left[\hat{\mathbf{f}}_{\mathrm{BLCU}}\hat{\mathbf{f}}_{\mathrm{BLCU}}'\right])^{-1/2}\mathrm{diag}(\epsilon\left[\hat{\mathbf{f}}_{\mathrm{DBLCP}}\hat{\mathbf{f}}_{\mathrm{DBLCP}}'\right])^{-1/2} = \mathbf{R}_{\mathrm{BLCU,DBLCP}} = \mathbf{I}.$$

*Proof.* The covariance between $\hat{\mathbf{f}}_{\mathrm{BLCU}}$ and $\hat{\mathbf{f}}_{\mathrm{DBLCP}}$ is

$$\mathbf{C}_{\mathrm{BLCU,DBLCP}} = (\mathbf{\Lambda}'\mathbf{\Sigma}^{-1}\mathbf{\Lambda})^{-1}\mathbf{\Lambda}'\mathbf{\Sigma}^{-1}\mathbf{xx}'\mathbf{\Sigma}^{-1}\mathbf{\Lambda}\mathbf{\Phi}^{1/2}(\mathbf{\Phi}^{1/2}\mathbf{\Lambda}'\mathbf{\Sigma}^{-1}\mathbf{\Lambda}\mathbf{\Phi}^{1/2})^{-1/2}\mathbf{\Phi}^{1/2}$$

$$= \mathbf{\Phi}^{1/2}(\mathbf{\Phi}^{1/2}\mathbf{\Lambda}'\mathbf{\Sigma}^{-1}\mathbf{\Lambda}\mathbf{\Phi}^{1/2})^{-1/2}\mathbf{\Phi}^{1/2}. \tag{9}$$

The corresponding correlation is

$$\mathbf{R}_{\mathrm{BLCU,DBLCP}} = \mathbf{\Phi}^{1/2}(\mathbf{\Phi}^{1/2}\mathbf{\Lambda}'\mathbf{\Sigma}^{-1}\mathbf{\Lambda}\mathbf{\Phi}^{1/2})^{-1/2}\mathbf{\Phi}^{1/2}\mathrm{diag}((\mathbf{\Lambda}'\mathbf{\Sigma}^{-1}\mathbf{\Lambda})^{-1})^{-1/2} \ . \tag{10}$$

If $\mathbf{\Phi} = \mathbf{I}$ and $\mathbf{\Lambda}'\mathbf{\Sigma}^{-1}\mathbf{\Lambda} = \mathrm{diag}(\mathbf{\Lambda}'\mathbf{\Sigma}^{-1}\mathbf{\Lambda})$ Equation 10 can be transformed into

$$\mathbf{R}_{\mathrm{BLCU,DBLCP}} = (\mathbf{\Lambda}'\mathbf{\Sigma}^{-1}\mathbf{\Lambda})^{-1/2}(\mathbf{\Lambda}'\mathbf{\Sigma}^{-1}\mathbf{\Lambda})^{1/2} = \mathbf{I} . \tag{11}$$

This completes the proof.                                                                                          □

To sum up, it has been shown that $\hat{\mathbf{f}}_{\mathrm{BL}}$, $\hat{\mathbf{f}}_{\mathrm{BLCU}}$, and $\hat{\mathbf{f}}_{\mathrm{DBLCP}}$ are perfectly correlated for corresponding orthogonal factors with $\mathbf{\Lambda}'\mathbf{\Sigma}^{-1}\mathbf{\Lambda} = \mathrm{diag}(\mathbf{\Lambda}'\mathbf{\Sigma}^{-1}\mathbf{\Lambda})$.

**Transformation resulting in perfect correlations between $\hat{\mathbf{f}}_{\mathrm{BL}}$, $\hat{\mathbf{f}}_{\mathrm{BLCU}}$, and $\hat{\mathbf{f}}_{\mathrm{DBLCP}}$**

The transformation of the factor loadings $\mathbf{\Lambda}$ into

$$\mathbf{\Lambda}^* = \mathbf{\Lambda}(\mathbf{\Lambda}'\mathbf{\Sigma}^{-1}\mathbf{\Lambda})^{-1/2}\mathrm{diag}(\mathbf{\Lambda}'\mathbf{\Sigma}^{-1}\mathbf{\Lambda})^{1/2} \tag{12}$$

yields

$$\mathbf{\Lambda}^{*'}\mathbf{\Sigma}^{-1}\mathbf{\Lambda}^* = \mathrm{diag}(\mathbf{\Lambda}'\mathbf{\Sigma}^{-1}\mathbf{\Lambda})^{1/2}(\mathbf{\Lambda}'\mathbf{\Sigma}^{-1}\mathbf{\Lambda})^{-1/2}\mathbf{\Lambda}'\mathbf{\Sigma}^{-1}\mathbf{\Lambda}(\mathbf{\Lambda}'\mathbf{\Sigma}^{-1}\mathbf{\Lambda})^{-1/2}\mathrm{diag}(\mathbf{\Lambda}'\mathbf{\Sigma}^{-1}\mathbf{\Lambda})^{1/2}$$

$$= \mathrm{diag}(\mathbf{\Lambda}'\mathbf{\Sigma}^{-1}\mathbf{\Lambda}). \tag{13}$$

This transformation modifies the factor inter-correlations as long as $\boldsymbol{\Lambda}'\boldsymbol{\Sigma}^{-1}\boldsymbol{\Lambda} \neq \mathbf{I}$, as follows from

$$\boldsymbol{\Phi} \neq \boldsymbol{\Phi}^{*}$$

$$(\boldsymbol{\Lambda}'\boldsymbol{\Lambda})^{-1}\boldsymbol{\Lambda}'(\boldsymbol{\Sigma}-\boldsymbol{\Psi}^{2})\boldsymbol{\Lambda}(\boldsymbol{\Lambda}'\boldsymbol{\Lambda})^{-1} \neq (\boldsymbol{\Lambda}^{*\prime}\boldsymbol{\Lambda}^{*})^{-1}\boldsymbol{\Lambda}^{*\prime}(\boldsymbol{\Sigma}-\boldsymbol{\Psi}^{2})\boldsymbol{\Lambda}^{*}(\boldsymbol{\Lambda}^{*\prime}\boldsymbol{\Lambda}^{*})^{-1}$$

$$(\boldsymbol{\Lambda}'\boldsymbol{\Lambda})^{-1}\boldsymbol{\Lambda}'(\boldsymbol{\Sigma}-\boldsymbol{\Psi}^{2})\boldsymbol{\Lambda}(\boldsymbol{\Lambda}'\boldsymbol{\Lambda})^{-1} \neq (\boldsymbol{\Lambda}'\boldsymbol{\Lambda}(\boldsymbol{\Lambda}'\boldsymbol{\Sigma}^{-1}\boldsymbol{\Lambda})^{-1/2}\operatorname{diag}(\boldsymbol{\Lambda}'\boldsymbol{\Sigma}^{-1}\boldsymbol{\Lambda})^{1/2})^{-1}\boldsymbol{\Lambda}'(\boldsymbol{\Sigma}-\boldsymbol{\Psi}^{2}) \tag{14}$$

$$\boldsymbol{\Lambda}(\operatorname{diag}(\boldsymbol{\Lambda}'\boldsymbol{\Sigma}^{-1}\boldsymbol{\Lambda})^{1/2}(\boldsymbol{\Lambda}'\boldsymbol{\Sigma}^{-1}\boldsymbol{\Lambda})^{1/2}\boldsymbol{\Lambda}'\boldsymbol{\Lambda})^{-1}.$$

Thus, even when the initial factor model was orthogonal ($\boldsymbol{\Phi}=\mathbf{I}$), the transformed factor model will not necessarily be orthogonal ($\boldsymbol{\Phi}^{*} \neq \mathbf{I}$). The transformation of the loadings according to Equation 12 can also be performed for correlated factors. According to Theorem 1, 2, and 3 it is, however, necessary to have orthogonal factors in order to get perfect correlations between $\hat{\mathbf{f}}_{BL}$, $\hat{\mathbf{f}}_{BLCU}$, and $\hat{\mathbf{f}}_{DBLCP}$ for corresponding factors. It is therefore proposed to perform a second order factor analysis so that

$$\boldsymbol{\Phi}^{*} = \boldsymbol{\Lambda}_{2}^{*}\boldsymbol{\Lambda}_{2}^{*\prime} + \boldsymbol{\Psi}_{2}^{2}, \tag{15}$$

where the subscript denotes the parameters of the second order factor model. A Schmid-Leiman (1957) transformation is than performed in order to compute orthogonal primary factors. It is possible to perform a Schmid-Leiman transformation of more complex hierarchical models. However, in purpose of brevity it is assumed here that $\boldsymbol{\Phi}^{*}$ can be decomposed into a single general (second order) factor and the corresponding uniqueness of the primary factors, that is

$$\boldsymbol{\Phi}^{*} = \boldsymbol{\Lambda}_{2}^{*}\boldsymbol{\Lambda}_{2}^{*\prime} + \boldsymbol{\Psi}_{2}^{*2} = \left[\boldsymbol{\Lambda}_{2}^{*} \vdots \boldsymbol{\Psi}_{2}^{*2}\right]\begin{bmatrix}\boldsymbol{\Lambda}_{2}^{*}\\\boldsymbol{\Psi}_{2}^{*2}\end{bmatrix} = \mathbf{P}\,\mathbf{P}'. \tag{16}$$

The Schmid-Leiman transformation of the oblique first order factor model is

$$\boldsymbol{\Lambda}_{SL}^{*} = \boldsymbol{\Lambda}^{*}\mathbf{P}. \tag{17}$$

It follows from Equations 2, 16, and 17 that

$$\boldsymbol{\Sigma} = \boldsymbol{\Lambda}'\boldsymbol{\Phi}^{*}\boldsymbol{\Lambda}^{*\prime} + \boldsymbol{\Psi}^{2} = \boldsymbol{\Lambda}_{SL}^{*}\boldsymbol{\Lambda}_{SL}^{*\prime} + \boldsymbol{\Psi}^{2}, \tag{18}$$

which implies that $\boldsymbol{\Lambda}_{SL}$ represents the loadings of orthogonal factors. In the most simple Schmid-Leiman solution, the first column in $\boldsymbol{\Lambda}_{SL}$ contains the loadings of the observed variables on a general (second order) factor that is orthogonal to the remaining orthogonalized primary factors. However, the interest here is into the orthogonalized primary factors which can be found in the columns 2 to $q$,

$$\boldsymbol{\Lambda}_{\mathrm{SLP}}^{*} = \begin{bmatrix} \lambda_{\mathrm{SL},1,2}^{*} & \cdots & \lambda_{\mathrm{SL},1,q}^{*} \\ \vdots & \ddots & \vdots \\ \lambda_{\mathrm{SL},p,2}^{*} & \cdots & \lambda_{\mathrm{SL},p,q}^{*} \end{bmatrix}. \tag{19}$$

The subset of orthogonalized primary factors can also be calculated by means of

$$\boldsymbol{\Lambda}_{\mathrm{SLP}}^{*} = \boldsymbol{\Psi}_{2}^{2} \boldsymbol{\Lambda}_{2}^{*}. \tag{20}$$

According to Equation 13 this implies

$$\boldsymbol{\Lambda}_{\mathrm{SLP}}^{*\prime} \boldsymbol{\Sigma}^{-1} \boldsymbol{\Lambda}_{\mathrm{SLP}}^{*} = \boldsymbol{\Psi}_{2}^{*2} \boldsymbol{\Lambda}^{*\prime} \boldsymbol{\Sigma}^{-1} \boldsymbol{\Lambda}^{*} \boldsymbol{\Psi}_{2}^{*2} = \boldsymbol{\Psi}_{2}^{*2} \mathrm{diag}(\boldsymbol{\Lambda}^{*} \boldsymbol{\Sigma}^{-1} \boldsymbol{\Lambda}) \boldsymbol{\Psi}_{2}^{*2} = \mathrm{diag}(\boldsymbol{\Lambda}_{\mathrm{SLP}}^{*\prime} \boldsymbol{\Sigma}^{-1} \boldsymbol{\Lambda}_{\mathrm{SLP}}^{*}), \quad (21)$$

so that the conditions for perfect correlations of $\hat{\mathbf{f}}_{\mathrm{BL}}$, $\hat{\mathbf{f}}_{\mathrm{BLCU}}$, and $\hat{\mathbf{f}}_{\mathrm{DBLCP}}$ are met for the corresponding orthogonalized primaries.

## Discussion

The present paper explores conditions for a perfect correlation between three types of factor score predictors: The regression predictor or best linear predictor, the conditionally unbiased best linear predictor, and the determinant best linear correlation-preserving predictor. A perfect correlation between these factor score predictors for corresponding factors implies that the choice between these factor score predictors does not matter and that each type of factor score predictor will have the virtues of the other. That is, the conditionally unbiased best linear predictor will also be the best linear predictor, the determinant best linear correlation-preserving predictor, will have the virtue to be conditionally unbiased predictor, etc.. Thus, the conditions of a perfect correlation between the three types of factor score predictors for corresponding factors might be of interest for applied researchers, who want to calculate score predictors combining the different advantages.

McDonald and Burr (1967) already found that three types of factor score predictors similar to the predictors investigated here are perfectly correlated for one factor models and for the unrotated canonical factor model. In addition to these conditions, it was shown here that for orthogonal factors with $\boldsymbol{\Lambda}^{\prime} \boldsymbol{\Sigma}^{-1} \boldsymbol{\Lambda} = \mathrm{diag}(\boldsymbol{\Lambda}^{\prime} \boldsymbol{\Sigma}^{-1} \boldsymbol{\Lambda})$ the three factor score predictors are perfectly correlated. A method for transforming a loading matrix according to this condition was proposed. The factors resulting from this transformation are not necessarily orthogonal. However, it has been shown that the factors corresponding to $\boldsymbol{\Lambda}^{\prime} \boldsymbol{\Sigma}^{-1} \boldsymbol{\Lambda} = \mathrm{diag}(\boldsymbol{\Lambda}^{\prime} \boldsymbol{\Sigma}^{-1} \boldsymbol{\Lambda})$ should be orthogonal in order to provide perfect correlations between the three types of factor score predictors for corresponding factors. Therefore, a hierarchical Schmid-Leiman based solution was proposed in order to provide orthogonal primary factors corresponding to the abovementioned conditions.